\documentclass[aps,prl,reprint,groupedaddress,showpacs]{revtex4-1}
\usepackage{amsmath}
\usepackage{amssymb}
\usepackage{graphicx}
\usepackage[colorlinks,bookmarks=false,citecolor=red,linkcolor=blue,urlcolor=blue]{hyperref}
\def\Hhat{\hat{H}}

\def\bhat{\hat{b}}
\def\nhat{\hat{n}}

\def\Sbf{\textbf{S}}

\def\r{{\bf r}}
\def\R{{\bf R}}

\def\SrCuBorate{SrCu$_{2}$(BO$_{3}$)$_{2}$}
\def\ketZero{|{\tt 0}\rangle}
\def\ketOne{|{\tt 1}\rangle}
\def\ketTwo{|{\tt 2}\rangle}
\def\braZero{\langle{\tt 0}|}
\def\braOne{\langle{\tt 1}|}
\def\braTwo{\langle{\tt 2}|}
\def\identity{\hat{\mathbf{1}}}
\def\htilde{\tilde{h}}
\def\calDhat{\hat{\mathcal{D}}}
\def\Vpprime{V^{\prime\prime}}
\def\Vppprime{V^{\prime\prime\prime}}
\begin{document}
\title{Effective models for the magnetization behavior of Shastry-Sutherland model}
\author{Brijesh Kumar} \email{bkumar@mail.jnu.ac.in}
\author{Bimla Danu}
\affiliation{School of Physical Sciences, Jawaharlal Nehru University, New Delhi 110067, India}
%\date{June 17, 2013} 
\date{\today}
\begin{abstract}  
The minimal effective model for the magnetisation plateaus below 1/2 in the Shastry-Sutherland model is derived to be an Ising model of certain unit-cell hard-core bosons with anisotropic repulsive interactions on isosceles triangular lattice. It unambiguously gives the prominent plateaus at 1/8, 1/6, 1/4, 1/3 and an additional one at 3/8, related through a particle-hole (p-h) transformation. The Dzyaloshinsky-Moriya (DM) interaction dresses it up with an inhomogeneous transverse field that also causes p-h asymmetry due to an in-plane component of the inter-dimer DM vector. This explains the asymmetry between the magnetisation below and above 1/4 in \SrCuBorate. The effective model above 1/2 plateau is an XXZ model with stronger XY parts. It gives no magnetisation plateaus, but exhibits chiral order.
\end{abstract}
\maketitle

%%%% 
The Shastry-Sutherland (SS) model, and the material \SrCuBorate~that realises it, are subjects of great current interest~\cite{Shastry1981, Kageyama1999}. This compound is a layered spin-gapped Mott insulator in which the Cu$^{2+}$ dimers  in CuBO$_{3}$ layers form the frustrated SS lattice of antiferromagnetically coupled quantum spin-1/2's~\cite{Miyahara1999,Shastry.Kumar}. The most notable feature of \SrCuBorate~is the occurrence of plateaus in the magnetisation, $M$, as a function of the magnetic field, $h$, at certain fractional values of the saturation magnetisation, $M_{sat}$~\cite{Kageyama1999,Onizuka2000}. This phenomenon has drawn much attention, and inspired a lot of studies.

Experimentally, the most prominent plateaus occur at $M/M_{sat}=1/8$, $1/4$ and $1/3$. The other plateaus at $1/9$, $1/7$, $1/6$, $1/5$ and $2/9$ have also been reported through torque measurements~\cite{Suchitra2008}. The plateau at $1/6$ has been confirmed recently, and an additional one at $2/15$ has been reported~\cite{Takigawa2013}. Above $1/3$, the plateaus at $2/5$ and $1/2$ have also been reported~\cite{Jaime2012}. While the plateau at $1/2$ has been confirmed, the one at $2/5$ seems absent, in recent ultra-high field measurements upto 118T~\cite{Matsuda2013}. Except $1/8$, $1/6$, $1/4$, $1/3$ and $1/2$, there appears to be a lack of consensus on the more exotic fractions. 

\begin{figure}[t]
   \centering
   \includegraphics[width=.48\textwidth]{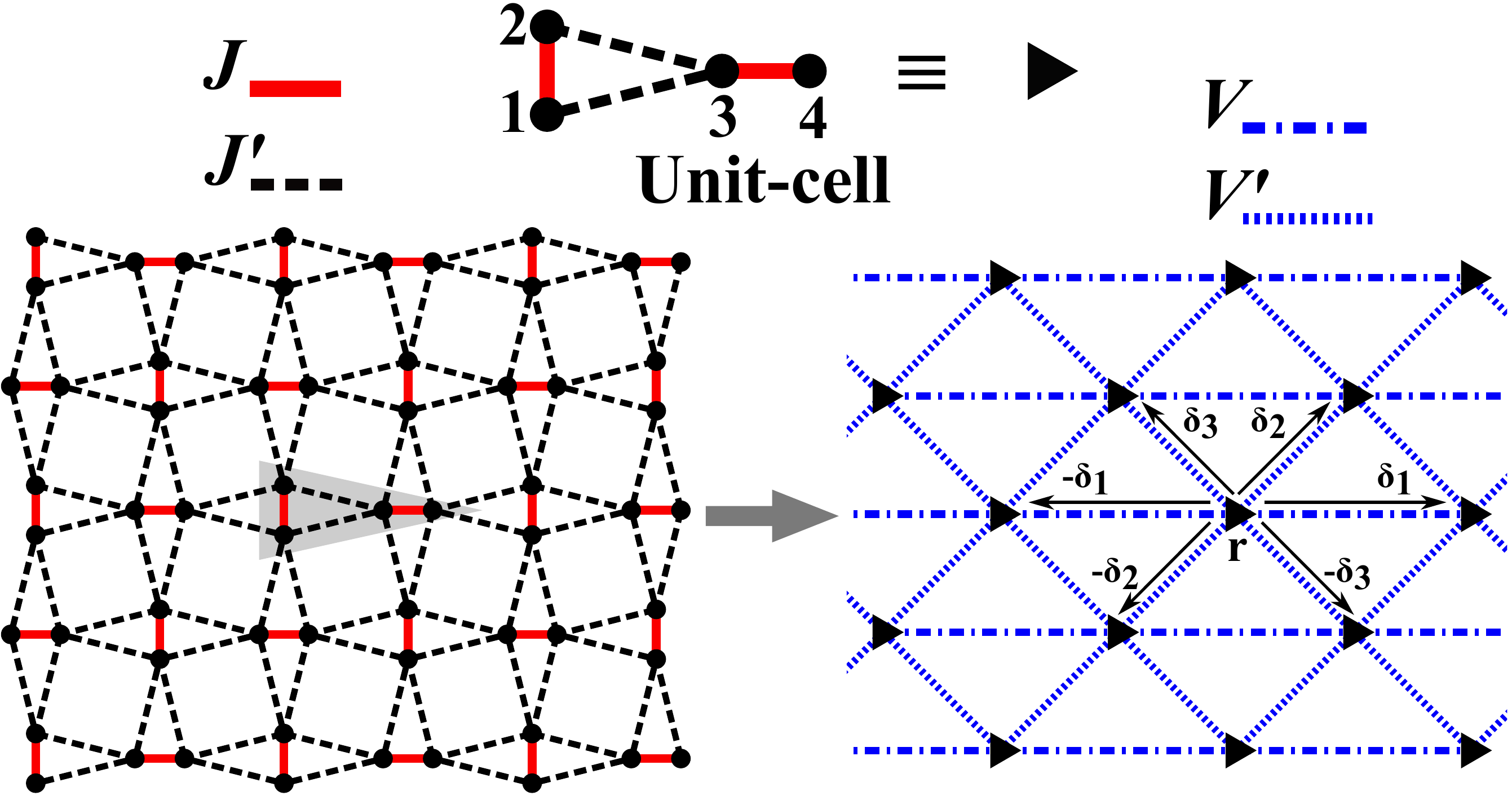} 
   \caption{The Shastry-Sutherland model Eq.~(\ref{eq:SS}), and the corresponding isosceles triangular lattice model Eq.~(\ref{eq:H0}).}
   \label{fig:SS_tri}
\end{figure}

These discoveries have led to a great deal of research on the Shastry-Sutherland model, 
$\Hhat_{SS}$, which is the basic quantum spin-1/2 model for \SrCuBorate~\cite{Miyahara.Review}.
\begin{equation}
\Hhat_{SS} = J\sum_{\langle i,j\rangle} \Sbf_{i}\cdot\Sbf_{j} + J^{\prime}\sum_{\langle l,m\rangle}  \Sbf_{l}\cdot\Sbf_{m} - h\sum_{i}S^{z}_{i}
\label{eq:SS}
\end{equation}
Here, $J$ is the intra-dimer exchange and $J^{\prime}$ is the inter-dimer coupling, both antiferromagnetic (see Fig.~\ref{fig:SS_tri}). For \SrCuBorate, various estimates give $J^{\prime}/J \approx 0.63$~\cite{Miyahara.Review,Matsuda2013}, which implies that its low temperature spin-gapped phase is a direct-product of the singlets on Cu$^{2+}$ dimers (the exact ground state of $\Hhat_{SS}$)~\footnote{The dimer-singlet state is the exact ground state of $H_{SS}$ for $J^{\prime}/J \lesssim 0.677$~\cite{Koga2000,Corboz2013}.}. The Dzyaloshinsky-Moriya (DM) interaction is also present as a small perturbation to the leading picture set by the SS model~\cite{DM.Cepas2001,DM.Kodama2005,DM.Penc2011}.

The early studies on $\Hhat_{SS}$ found the dimer-triplet excitations to be highly localised objects~\cite{Miyahara1999}. Thus, for $h\neq 0$, the magnetisation behaviour of the SS model, and that of the \SrCuBorate, is understood to be dominated by the interactions between the field-induced triplets (effective hard-core bosons), competed at best by the correlated hopping processes~\cite{,Miyahara.Review, Momoi2000}. Accordingly, the crystalline superstructures of the localised triplets, stabilised by interactions, characterise the magnetisation plateaus, as observed at $1/8$ through NMR~\cite{Kodama2002}. Away from a plateau, but near its ends, there could also arise a supersolid phase, wherein $M$ grows smoothly, while the crystalline order of triplets has not melted~\cite{Momoi2000,Takigawa2008}. These ideas have advanced to sophisticated levels of computation through the pCUT (perturbative continuous unitary transformations)~\cite{Dorier2008} and CORE (contractor-renormalization) methods~\cite{Abendschein2008}. An alternate approach is the Chern-Simons (CS) theory~\cite{Misguich2001}, which has predicted a series of plateaus at $1/q$ for $9\ge q \ge 2$ and $2/9$~\cite{Suchitra2008}. 
 
While the effective models in terms of the dimer-hard-core bosons~\footnote{By a dimer-hard-core boson, we mean the usual two-level object given by the singlet and a fully polarised triplet on a dimer of quantum spins [Cu$^{2+}$ ions in \SrCuBorate]. In a magnetic field, these two dimer states come together to form a minimal dimer-subspace which is typically considered relevant to discuss the magnetisation properties of the dimerised antiferromagnets.} have been generated to very high orders in $J^{\prime}/J$, and they do give a host of plateaus, but they look obscure and don't offer much clarity into whether a plateau occurs and why.  
In this Letter, we try to change this situation by deriving a simple effective model for the magnetisation behaviour of the SS model. Through this, we unambiguously get all the prominent plateaus at $1/8$, $1/6$, $1/4$, $1/3$ and $1/2$, and an additional one at $3/8$. We also understand that the plateaus at $1/8$ and $3/8$ occur as a pair, and the same for $1/6$ and $1/3$, related via the particle-hole transformation in the minimal model, and how DM interaction affects this feature. Our effective model for $M/M_{sat}\le 1/2$, in its barest form, is a classical problem (Ising model) of hard-core bosons with repulsive interactions on isosceles triangular lattice, which is dressed by the transverse-field like quantum fluctuations due to DM interaction.  
We also derive a minimal effective model for $M/M_{sat}\ge 1/2$, which is an XXZ problem on isosceles triangular lattice. It gives no plateaus (except some anomalies near $1/2$ and 1), but a chiral order.  

 %%%%%%%%%%%%%%%%
The basic premise of our study is a doubt whether the dimer-hard-core bosons can ever lead to a neat and decisive understanding of the magnetisation behaviour of the SS  model~\footnote{For instance, a similar lack of faith in the standard belief that dimer-triplets crystallise to form plateaus in SS model has been expressed recently in Ref.~\cite{Philippe2014}.}.
It is because the orthogonal-dimer topology of the SS lattice renders the $J^{\prime}$ links around a dimer-singlet 
ineffective by annihilation, which makes it hard to reconstruct the dynamics in terms of the dimer-hard-core bosons. It comes as no surprise, therefore, that the corresponding effective models remain obscure, and despite the progress, leave one in doubt about the finality of their outcome. 

\begin{figure}[b]
\centering
\includegraphics[width=0.237\textwidth]{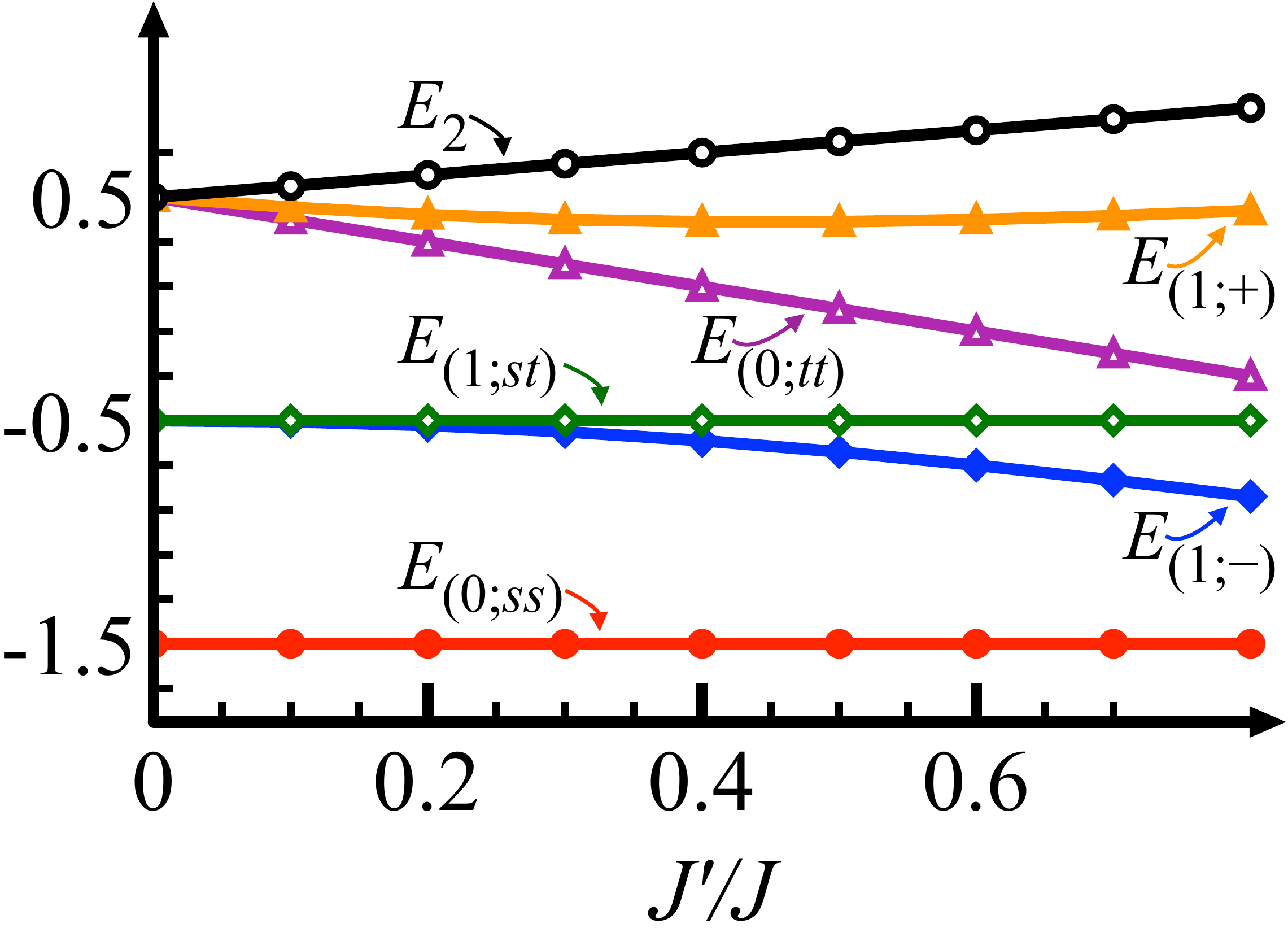} \hfill 
\includegraphics[width=0.24\textwidth]{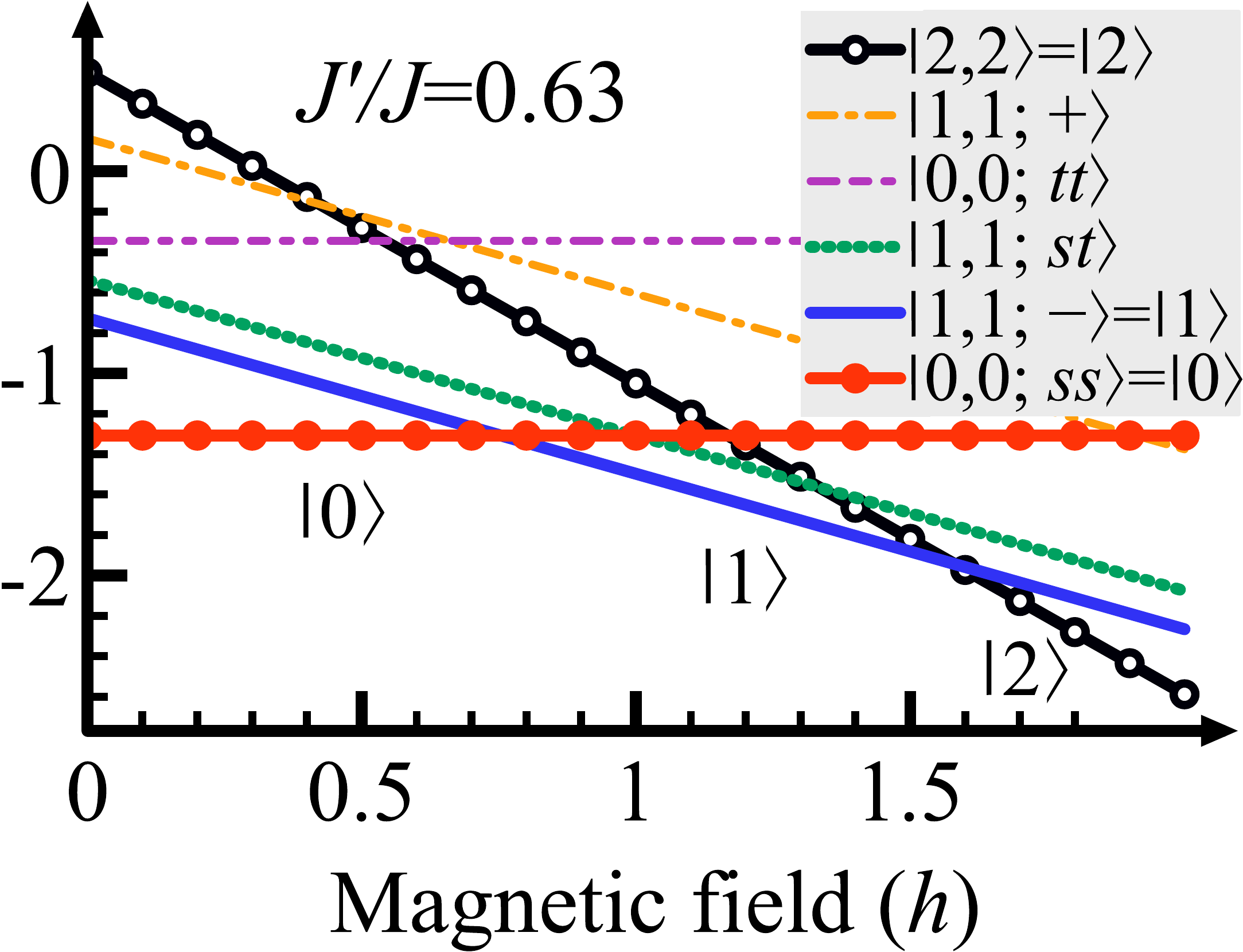}
\caption{Eigenvalues of a single unit-cell of the Shastry-Sutherland model plotted as a function of  $J^{\prime}/J$, and $h$.}
\label{fig:cell-eigenvalues}
\end{figure}

We overcome this difficulty by working with the eigenstates of $J(\Sbf_{1}\cdot\Sbf_{2} + \Sbf_{3}\cdot\Sbf_{4})+J^{\prime}(\Sbf_{1}+\Sbf_{2})\cdot\Sbf_{3}$, a natural `crystallographic' unit-cell of the SS lattice (see Fig.~\ref{fig:SS_tri}). These states carry in them at least some effect of the $J^{\prime}$ exactly, which for the dimer-states would require much extra effort to reconstruct. These eigenstates are presented in Table~\ref{tab:cell-states}, and their eigenvalues are plotted as a function of  $J^{\prime}/J$, and $h$, in Fig.~\ref{fig:cell-eigenvalues}. For a basic effective description of the SS model in magnetic field, it would suffice to work with $\ketZero \equiv |0,0;ss\rangle$ and $\ketOne \equiv |1,1;-\rangle$ for $M/M_{s}\le \frac{1}{2}$, and $\ketOne$ and $\ketTwo\equiv |2,2\rangle$ for $M/M_{sat}\ge \frac{1}{2}$.

\begin{figure}[b]
\centering
\includegraphics[width=0.238\textwidth]{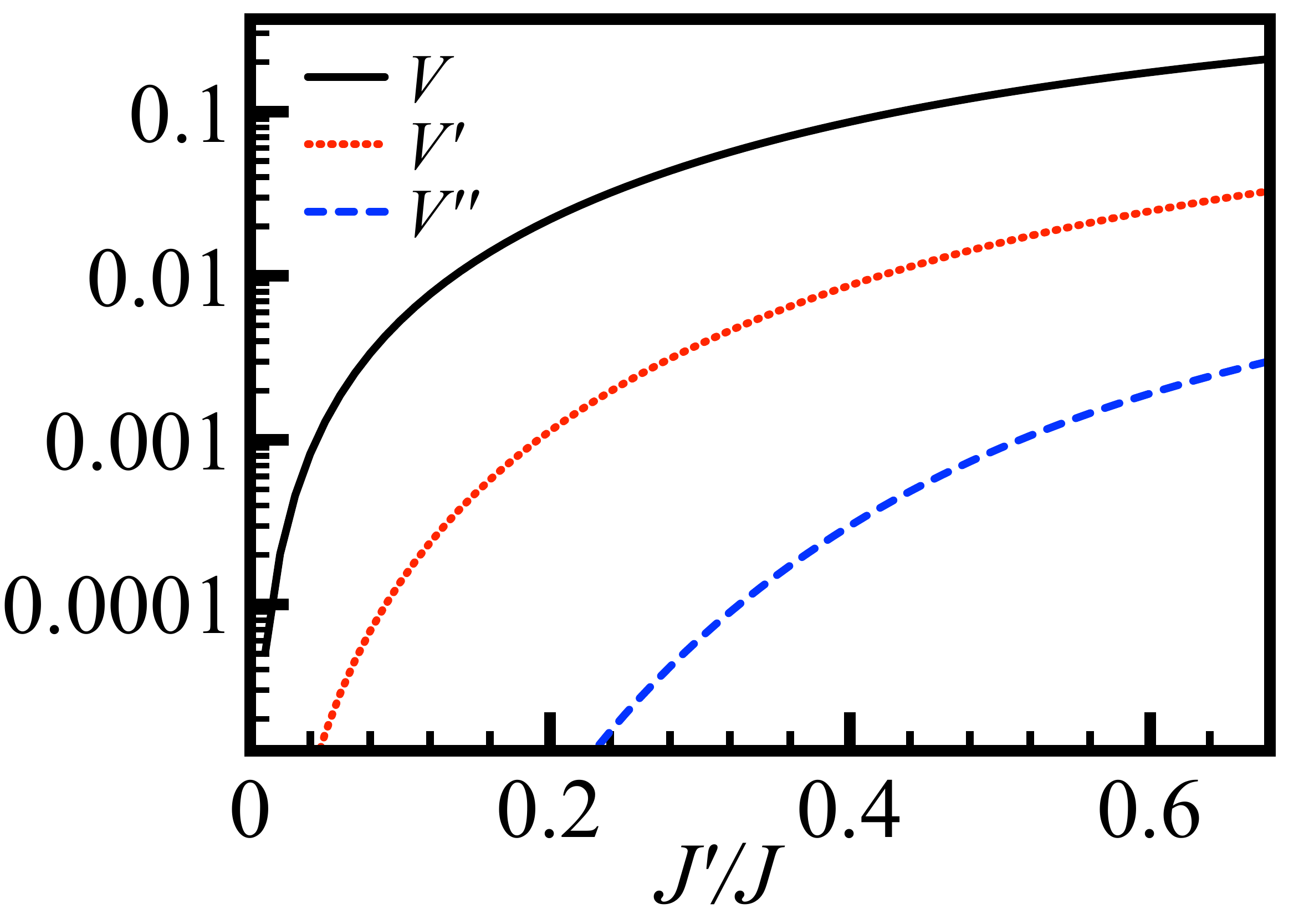} 
\includegraphics[width=0.238\textwidth]{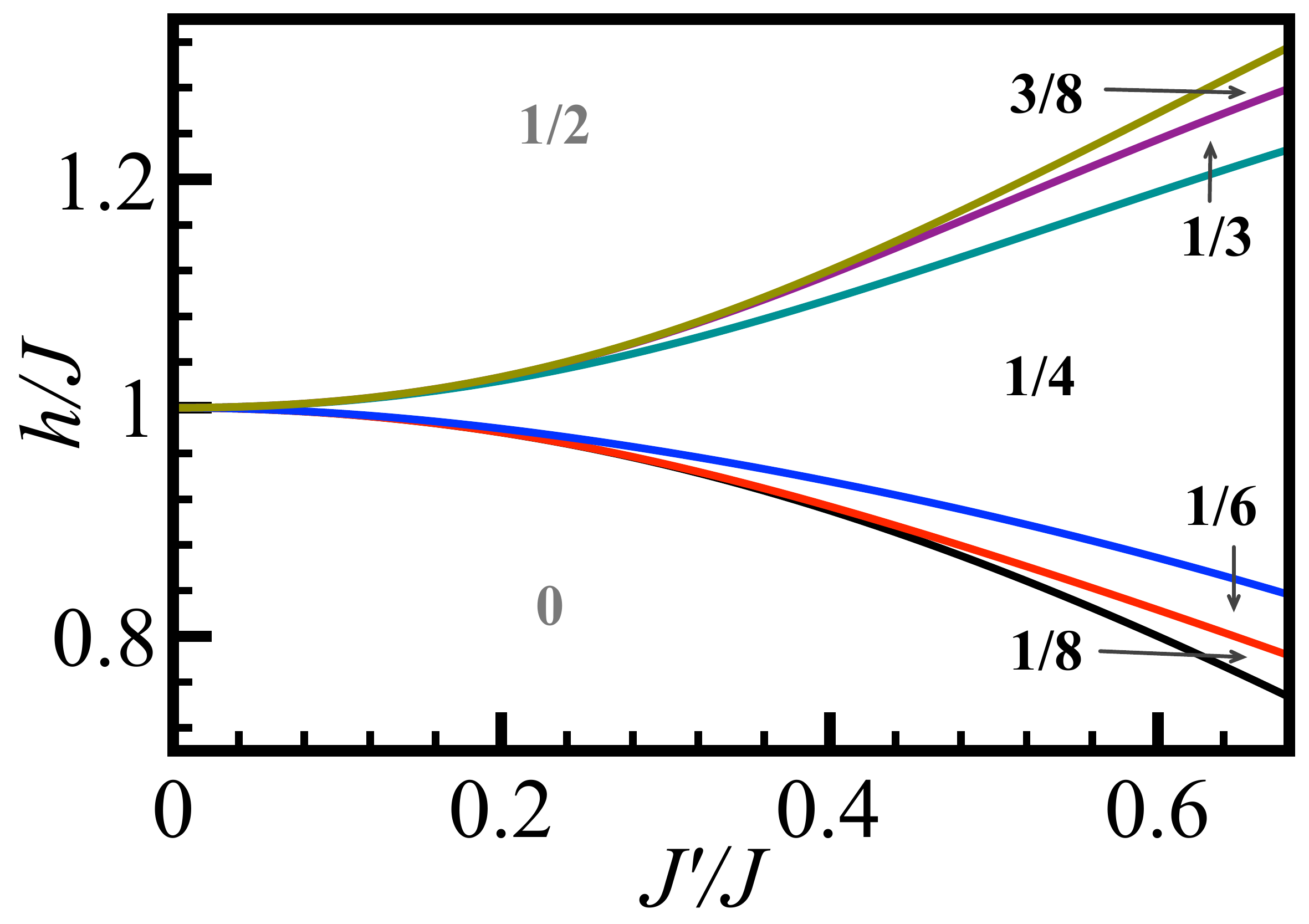}
\caption{The interactions (left), and the plateau phase diagram (right), within $\Hhat_{0}+\Hhat_{X}$ for $0\le M/M_{sat}\le 1/2$.} 
\label{fig:Ising_Xchange}
\end{figure}

For $M/M_{sat}\le 1/2$, we derive the effective Hamiltonian, $\Hhat_{\le\frac{1}{2}} = E_{(0;ss)}L+\Hhat_{0}+\Hhat_{X}  + \Hhat_{DM} $, by reorganising the SS lattice in terms of these unit-cells without breaking the translational symmetry, and projecting each unit-cell onto its $\{\ketZero$, $\ketOne\}$ basis. Here, $E_{(0;ss)}L$ is the singlet energy of $L$ unit-cells. The $\Hhat_{0}$ is the \emph{minimal} effective Hamiltonian for the SS model, in terms of the ``cell-hard-core bosons''  (defined in Table~\ref{tab:Rep01}), on the isosceles triangular lattice of Fig.~\ref{fig:SS_tri}, with total $L$ sites.
\begin{equation}
\Hhat_{0} = \sum_{\r}\nhat_{\r}\left[-\htilde+V\nhat_{\r+\delta_{1}}+V^{\prime}
\left(\nhat_{\r+\delta_{2}}+\nhat_{\r+\delta_{3}}\right)\right]
  \label{eq:H0}
  \end{equation}
The effective interactions, $V=J^\prime(3+\cos{\theta})(1-\cos{\theta}+2\sqrt{2}\sin{\theta})/32$ and $V^\prime=J^\prime(3+\cos{\theta})(1-\cos{\theta})/32$, are repulsive, and $V $ is always stronger than $V^\prime$ (see Fig.~\ref{fig:Ising_Xchange}). The `chemical potential', $\htilde=h-\Delta_{1}$, controls the filling of these hard-core particles with energy-gap $\Delta_{1}=E_{(1;-)}-E_{(0;ss)}$.  

\begin{table}[b]
\caption{\label{tab:cell-states} The eigenstates of a single unit-cell of the Shastry-Sutherland model (see Fig.~\ref{fig:SS_tri}), in terms of the singlet, $|s\rangle$, and the triplets, $|t_{m}\rangle$ for $m=1,0,\bar{1}$, on the bonds (1,2) and (3,4).\footnote{The bond-states on (1,2) are: $|s\rangle_{12} = \frac{|\uparrow_1\downarrow_2\rangle - |\downarrow_1\uparrow_2\rangle}{\sqrt{2}}$, $|t_1\rangle_{12} = |\uparrow_1\uparrow_2\rangle$, $|t_0\rangle_{12} = \frac{|\uparrow_1\downarrow_2\rangle + |\downarrow_1\uparrow_2\rangle}{\sqrt{2}}$ and $|t_{\bar 1}\rangle_{12} = |\downarrow_1\downarrow_2\rangle$, and likewise on bond (3,4). The negative $m$'s are denoted as $\bar{m}$.}}
\begin{ruledtabular}
\begin{tabular}{p{.47\textwidth}}
Eigenstates~\footnote{The eigenstates are denoted as $|S_{uc},m_{uc}; \mbox{\footnotesize extra-labels}\rangle$, where $S_{uc}$ is the total spin quantum-number of the unit-cell, $m_{uc}$ is the corresponding $z$-component, and the `extra-labels' indicate (when necessary) the spins of the bond-states [on (1,2) and (3,4), respectively] that make it.} \hfill Eigenvalues  \\ \colrule
\textbf{Singlets} \\ 
 $|0,0;ss\rangle =|s\rangle_{12} |s\rangle_{34}$ \hfill $E_{(0;ss)} = -3J/2$ 
\\
  $|0,0;tt\rangle = \frac{|t_1\rangle_{12} |t_{\bar 1}\rangle_{34} - |t_0\rangle_{12} |t_0\rangle_{34} +  |t_{\bar 1}\rangle_{12} |t_{1}\rangle_{34}}{\sqrt{3}}$ \hfill $E_{(0;tt)} = \frac{J}{2}-J^\prime$ 
   \\ 
 \textbf{Triplets}\footnote{Here, the triplet states $|1,m;ts\rangle = |t_m\rangle_{12}|s\rangle_{34}$, and $|1,m;tt\rangle$ are given as: $|1,1;tt\rangle =\{|t_1\rangle_{12}|t_0\rangle_{34}-|t_0\rangle_{12}|t_1\rangle_{34}\}/\sqrt{2}$, $|1,0;tt\rangle =\{|t_1\rangle_{12}|t_{\bar 1}\rangle_{34}-|t_{\bar 1}\rangle_{12}|t_1\rangle_{34}\}/\sqrt{2}$, and $|1,{\bar 1};tt\rangle =\{|t_0\rangle_{12}|t_{\bar 1}\rangle_{34}-|t_{\bar 1}\rangle_{12}|t_0\rangle_{34}\}/\sqrt{2}$. Moreover, $\theta_{\pm}=\theta+\pi^{\frac{1\mp1}{2}}$, where $\tan{\theta}=2\sqrt{2}J^{\prime}/(2J-J^{\prime})$.}
\\ 
 $ |1,m;st\rangle =|s\rangle_{12} |t_{m}\rangle_{34}$ \hfill $E_{(1;st)} = -J/2$ 
 \\ 
  $|1,m;\pm\rangle =\cos{(\theta_{\pm}/2)}|1,m;tt\rangle  + \sin{(\theta_{\pm}/2)}|1,m;ts\rangle $ \\ 
\hfill $E_{(1;\pm)} = -\frac{J^\prime}{4}\pm\sqrt{\left(\frac{J}{2}-\frac{J^\prime}{4}\right)^2+\frac{{J^\prime}^2}{2}} $
 \\ 
\textbf{Quintets}~\footnote{The states $|2,{\bar 2}\rangle$ and $|2,{\bar 1}\rangle$ can be obtained from $|2,2\rangle$ and $|2,1\rangle$, respectively, by replacing $|t_{1}\rangle$ with $|t_{\bar 1}\rangle$.} \\
$|2,2\rangle = |t_1\rangle_{12}|t_1\rangle_{34}$ \hfill $E_{2}  = (J+J^{\prime})/2$ \\
$|2,1\rangle  = \frac{|t_1\rangle_{12}|t_0\rangle_{34} + |t_0\rangle_{12}|t_1\rangle_{34}}{\sqrt{2}}$ \\
$ |2,0\rangle = \frac{|t_1\rangle_{12}|t_{\bar 1}\rangle_{34} + 2|t_0\rangle_{12}|t_0\rangle_{34} + |t_0\rangle_{12}|t_1\rangle_{34}}{\sqrt{6}}$ \\
\end{tabular}
\end{ruledtabular}
\end{table}

The $\Hhat_{X}$ denotes the corrections beyond $\Hhat_{0}$. One could generate it, say, by using the pCUT method with our unit-cell states. But here we do something very simple. We add some more repulsion, just a little beyond $V$ and $V^{\prime}$. That is, we consider $\Hhat_{X} = \sum_{\r}\nhat_{\r}[\Vpprime\nhat_{\r+\delta_{2}+\delta_{3}}+\Vppprime(\nhat_{\r+\delta_{1}+\delta_{2}}+\nhat_{\r-\delta_{1}+\delta_{3}})]$. This is not quite ad hoc. We know these interactions would be there, and expect $\Vppprime \lesssim \Vpprime$ to be very small~\footnote{Our $\Vpprime$ is similar to $V^{\prime}_{3}$ in Refs.~\cite{Dorier2008,Abendschein2008}, and $\Vppprime$ is roughly like their $V_{7}$ which is $\lesssim V^{\prime}_{3}$. While mostly $V^{\prime}_{3}$ and $V_{7}$ are negligible (consistent with the absence of $V^{\prime\prime}$ and $\Vppprime$ in our minimal $\Hhat_{0}$), but they begin to show up for big enough $J^{\prime}/J$. Our heuristic estimate of $V^{\prime\prime}$ is numerically not badly off from the values of $V^{\prime}_{3}$ in these references.}. Heuristically, we estimate $V^{\prime\prime} \sim \frac{V}{32}(\sin{\theta})^{4}$, and take $\Vppprime\approx \Vpprime$ for the calculations below.

\begin{table}
\caption{\label{tab:Rep01} Representation of the spins in a unit-cell in the basis, $\{\ketZero, \ketOne\}$, where $\ketZero = |0,0;ss\rangle$ and $\ketOne=|1,1;-\rangle$.\footnote{Here, $\nhat=\ketOne\braOne \equiv (\identity+\tau^{z})/2$, $\identity=\ketZero\braZero + \ketOne\braOne$ is the identity, $\tau^{+}=\ketOne\braZero \equiv \bhat^{\dag}$, $\tau^{x}=\tau^{+}+\tau^{-}$ and $\tau^{y}=-i(\tau^{+}-\tau^{-})$.}}
\begin{ruledtabular}
\begin{tabular}{p{.225\textwidth} | p{.23\textwidth}}
$S_{1x} = -S_{2x} = \frac{\cos{(\theta/2)}}{2\sqrt{2}} \tau^x$ & 
$S_{3x} = S_{4x} = S_{3y} = S_{4y} = 0$ \\ 
$S_{1y} = -S_{2y}= -\frac{\cos{(\theta/2)}}{2\sqrt{2}} \tau^y$ & 
$S_{3z}=\frac{(1-\cos{\theta}-2\sqrt{2}\sin{\theta})}{8}\nhat$ \\
$S_{1z} = S_{2z}= \frac{(3+\cos{\theta})}{8}\nhat$ & 
$S_{4z}=\frac{(1-\cos{\theta}+2\sqrt{2}\sin{\theta})}{8}\nhat$ \\
\end{tabular}
\end{ruledtabular}
\end{table}

\begin{figure}[t]
\centering
\includegraphics[width=.48\textwidth]{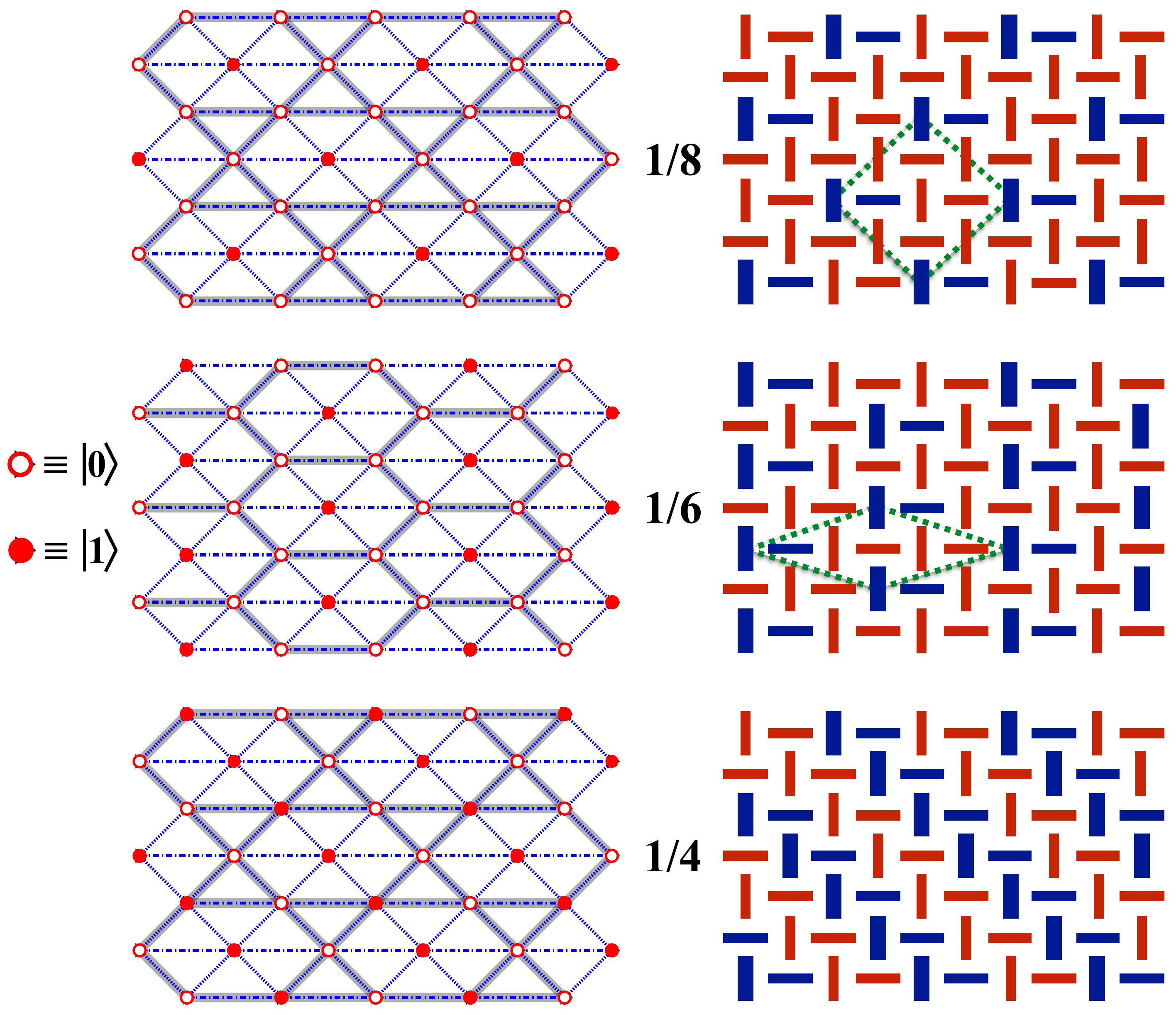}
\caption{The cell-triplet superlattices at $M/M_{sat}=1/8$, $1/6$ and $1/4$ plateaus within $\Hhat_{0}+\Hhat_{X}$, and the corresponding dimer-triplet orders on SS lattice. Here, the blue bonds with two different thicknesses denote dimer-triplets with different weights, and the red ones are the dimer-singlets. The ordered structures at $1/3$ and $3/8$ can be obtained by particle-hole transformation, $\ketZero\leftrightarrow\ketOne$, on $1/6$ and $1/8$ states, respectively.}
\label{fig:order1by864}
\end{figure}

Now we discuss $M/M_{sat}$ vs. $h$ using $\Hhat_{0}+\Hhat_{X}$. Here, $M=\sum_{\bf r} \langle\nhat_{\bf r}\rangle$ and $M_{sat}=2L$. While $h$ likes to populate the lattice with hard-core bosons (cell-triplets), the repulsive interactions like them to stay as far away from each other as possible. The first plateau would thus correspond to a filling that barely avoids repulsion. For $\Hhat_{0}$, it is $M/M_{sat}=1/6$ with a rhombic superlattice of hard-core particles (a honeycomb of `holes') shown in Fig.~\ref{fig:order1by864}. But a non-zero $V^{\prime\prime}$ or $\Vppprime$, \emph{howsoever small}, immediately realises $1/8$ as the lowest plateau with a square superlattice of cell-triplets (Kagom\'e lattice of holes) by pushing $1/6$ higher up in energy. The lower ($c1$) and upper ($c2$) critical fields for these plateaus (at $T=0 K$) are: $\htilde^{(\frac{1}{8})}_{c1}= 0$ and $\htilde^{(\frac{1}{8})}_{c2}= \htilde^{(\frac{1}{6})}_{c1}= 4(V^{\prime\prime}+2\Vppprime)$. The smallness of $\Vpprime$ and $\Vppprime$ is qualitatively consistent with the small experimental width ($\sim 1$ T) of $1/8$ plateau.

To deduce the higher $M$ plateaus, we use the particle-hole (p-h) transformation, $\nhat_{\r}\rightarrow \identity - \nhat_{\r}$, under which $M/M_{sat}$ $\rightarrow$ $\frac{1}{2}-M/M_{sat}$, 
$\htilde \rightarrow 2(V+2V^{\prime}+V^{\prime\prime}+2\Vppprime)-\htilde$, while $V$, $V^{\prime}$, $V^{\prime\prime}$ and $\Vppprime$ remain the same. Clearly, it implies a plateau at $3/8$ due to the one at $1/8$. Likewise, it gives a plateau at $1/3$ due to $1/6$. Since $\Hhat_{0}+\Hhat_{X}$ is invariant under p-h transformation for $\htilde= V+2V^{\prime}+V^{\prime\prime}+2\Vppprime$ at $M/M_{sat}=1/4$, it naturally brings in the plateau at $1/4$. Besides, the $1/2$ is trivially there. The superlattice structures at $1/3$ and $3/8$ plateaus are obtained by p-h transforming (${\tt 0}\leftrightarrow {\tt 1}$) the structures at $1/6$ and $1/8$, respectively. At $1/4$, the cell-triplets form stripes, as in Fig.~\ref{fig:order1by864}, consistent with what is known. The critical fields obtained by comparing the energies of these ordered states are: $\htilde^{(\frac{1}{6})}_{c2}= \htilde^{(\frac{1}{4})}_{c1}= 3V^{\prime}-2V^{\prime\prime}-\Vppprime$, and $\htilde^{(\frac{1}{4})}_{c2}=\htilde^{(\frac{1}{3})}_{c1}$, $\htilde^{(\frac{1}{3})}_{c2}=\htilde^{\frac{3}{8}}_{c1}$ and $\htilde^{(\frac{3}{8})}_{c2}=\htilde^{(\frac{1}{2})}_{c1}$ can be determined from the p-h transformation rule for $\htilde$. Thus, all the prominent plateaus ($1/8$, $1/6$, $1/4$, $1/3$ and $1/2$) arise naturally and unambiguously in our very simple effective model, $\Hhat_{0}+\Hhat_{X}$. It also gives an additional plateau at $3/8$. These we have also checked by exact energy minimisation (on small clusters) and monte carlo simulations of this effective model~\cite{Bimla.thesis}. Next we discuss the effects of DM interaction, relevant to \SrCuBorate, on $\Hhat_{0}+\Hhat_{X}$.

\begin{figure}[t]
\centering
\includegraphics[width=.238\textwidth]{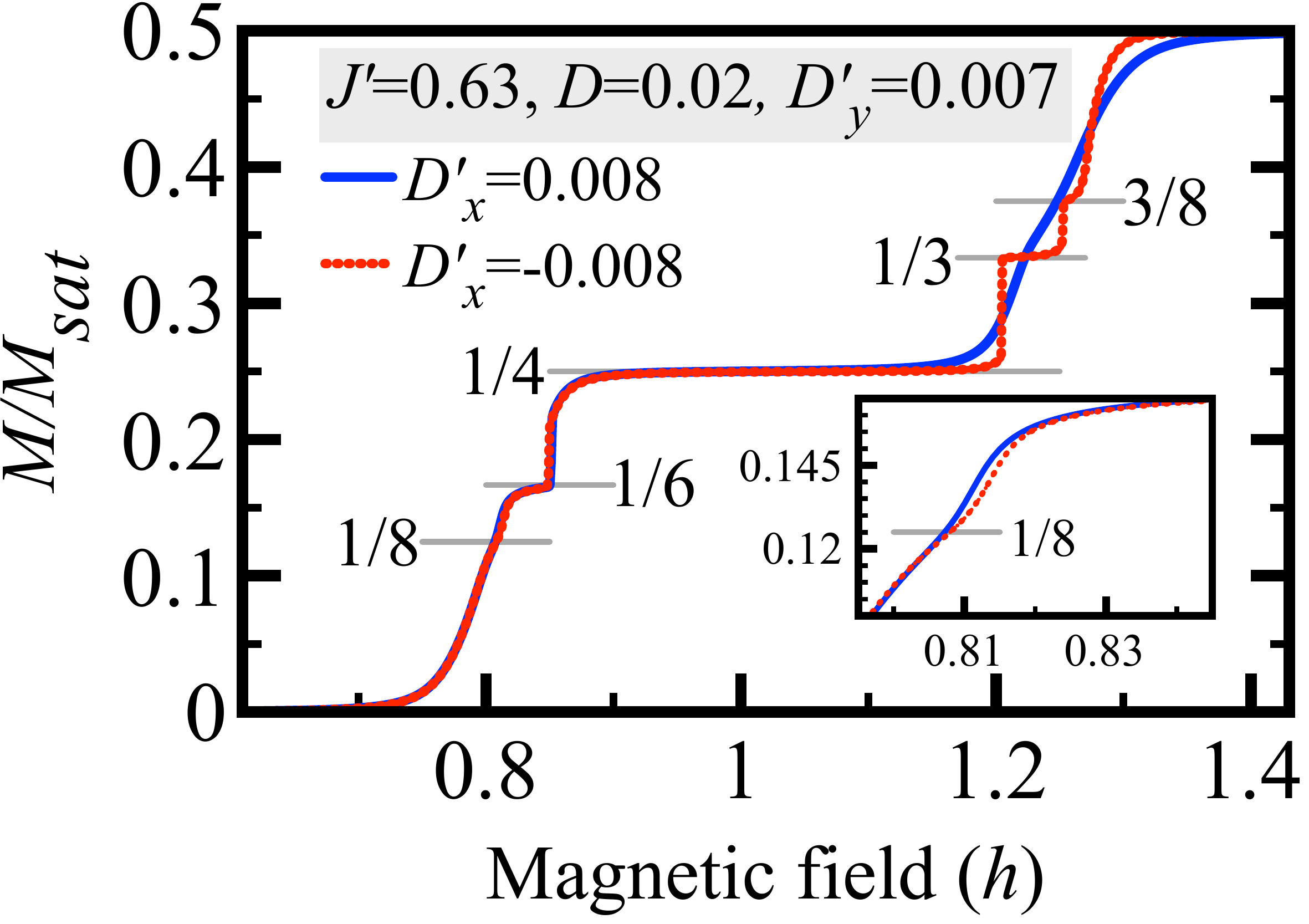} \hfill
\includegraphics[width=.238\textwidth]{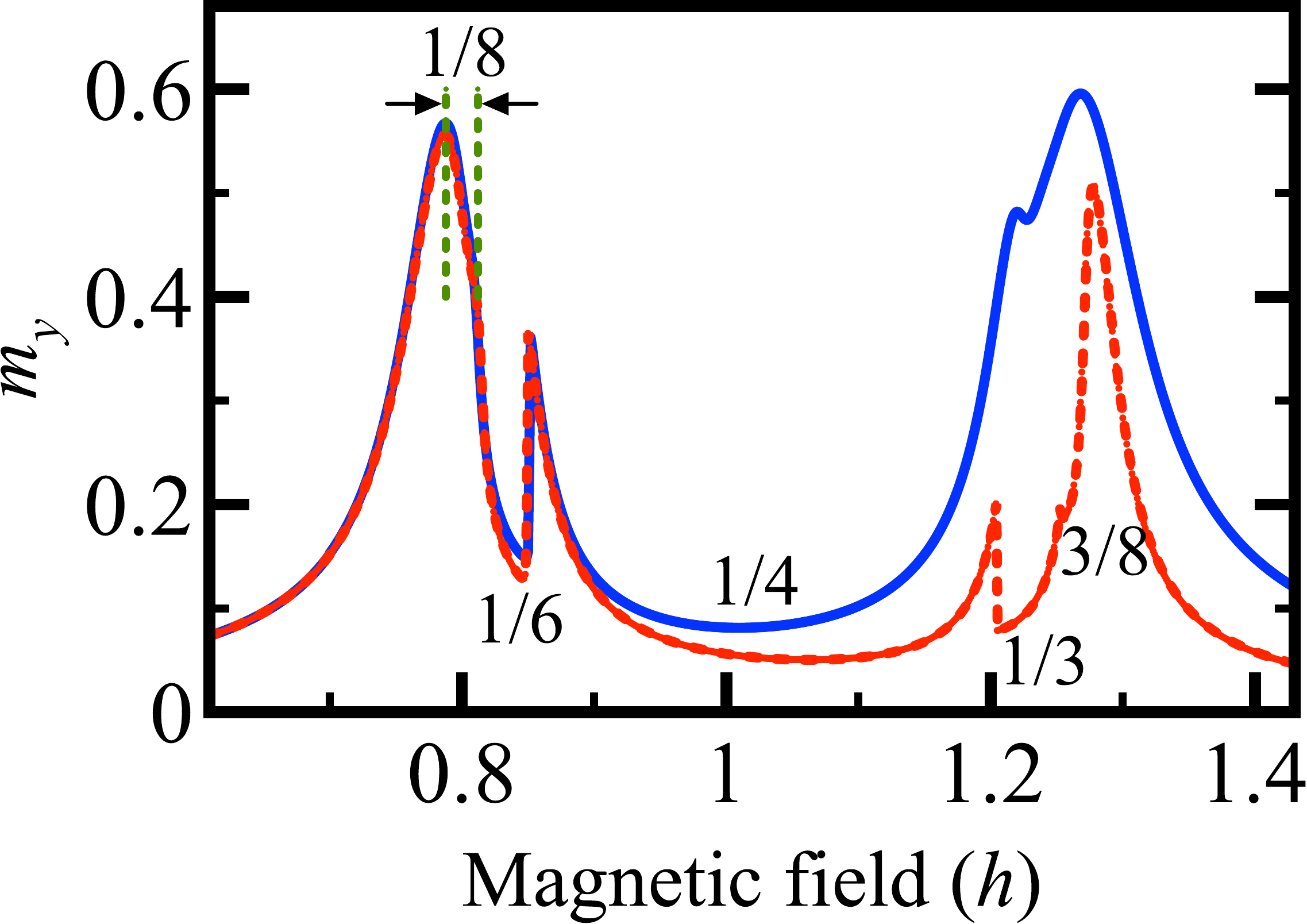}
\caption{(Left) $M$ vs. $h$ in the ground state of $\Hhat_{\le\frac{1}{2}}$. Notice the asymmetry below and above 1/4, pronounced differently for different signs of $D^{\prime}_{x}$.  (Inset) Enlarged view of magnetisation around 1/8. (Right) The transverse magnetisation, $m_{y}=\frac{1}{L}\sum_{\r}\langle\tau^{y}\rangle$, for the same values of parameters (with $J=1$).}
\label{fig:MvsH_VVpV2pV3pD}
\end{figure}

At low temperatures, the intra-dimer DM vector in \SrCuBorate~lies in the plane of dimers and $\perp$ to the dimer's orientation, and the inter-dimer DM interaction is kind of arbitrary~\cite{DM.Penc2011}. By projecting them onto the local $\{\ketZero, \ketOne\}$ basis, we get the following effective $\Hhat_{DM}$. 
\begin{equation}
\Hhat_{DM} = \sum_{\r} \left[\calDhat^{x}_{\r}(-\delta_{2},\delta_{3})\tau^{x}_{\r}+\calDhat^{y}_{\r}(-\delta_{1},-\delta_{2},\delta_{3})\tau^{y}_{\r}\right]
\label{eq:HlessDM}
\end{equation}
Here, $\calDhat^{x}_{\r}(-\delta_{2},\delta_{3}) = D^{\prime}_{y}(\sin{\theta}\cos{\frac{\theta}{2}})(\nhat_{\r+\delta_{3}}-\nhat_{\r-\delta_{2}})/4$ and $\calDhat^{y}_{\r}(-\delta_{1},-\delta_{2},\delta_{3}) = -\frac{D}{2\sqrt{2}}\cos{\frac{\theta}{2}} + \frac{D_{x}^{\prime}}{8\sqrt{2}}(\sin{\theta}\sin{\frac{\theta}{2}})[\nhat_{\r+\delta_{3}} + \nhat_{\r-\delta_{2}} -2(\frac{V}{V^{\prime}})\nhat_{\r-\delta_{1}}]$ are the effective `transverse fields' dependent upon the local occupancies, $\nhat_{\r}$'s. The intra-dimer DM interaction is denoted as $D$ ($\sim 0.03 J$), $D^{\prime}_{x,y}$ ($\lesssim D$) are the $x$ and $y$ components of the inter-dimer DM vector, and $V/V^{\prime} = 1+2\sqrt{2}\cot{\frac{\theta}{2}}$. Thus, the minimal effective model for the magnetisation behaviour of \SrCuBorate, $\Hhat_{0}+\Hhat_{X}+\Hhat_{DM} = \Hhat_{\le \frac{1}{2}}$, is a `quantum Ising' model with `dynamically' inhomogeneous transverse fields~\footnote{By dynamic inhomogeneity of a transverse field, here we mean that it is determined dynamically by the hard-core particles' occupancies at different sites.}. Under the p-h transformation, $D^{\prime}_{x}\rightarrow-D^{\prime}_{x}$, $D^{\prime}_{y} \rightarrow D^{\prime}_{y}$ and $D\rightarrow D+\frac{D^{\prime}_{x}}{\sqrt{2}}\sin{\theta}$. An important physical implication of these rules is that a non-zero $D^{\prime}_{x}$ [amplified by $V/V^{\prime}$ ($\sim 7$ for $J^{\prime}/J=0.63$)] causes p-h \emph{asymmetry} between the related plateaus, which is indeed there in \SrCuBorate. For instance, the plateaus at $1/6$ and $1/3$ differ in widths, and $3/8$ is still not seen, while $1/8$ is too well known. Such p-h asymmetry can also arise due to the three-body interactions in $\Hhat_{X}$, but here, we have discussed only the most essential physical content of the \SrCuBorate, viz. SS model, problem below $1/2$. The results of an exact numerical computation in the ground state of $\Hhat_{\le \frac{1}{2}}$ on a 12-sites periodic cluster are shown in Fig.~\ref{fig:MvsH_VVpV2pV3pD}. The $m_{y}$ there gives the staggered transverse magnetisation on (vertical) dimers. The jumps between successive plateaus (and in $m_{y}$) is due to the change in the underlying crystalline order (see Fig.~\ref{fig:order1by864}).
Clearly, the $\Hhat_{\le \frac{1}{2}}$ is in broad qualitative agreement with the experiments, and can be improved quantitatively by  pCUTs, CORE or any other suitable methods.

Interestingly, the recently suggested devil's staircase in \SrCuBorate~\cite{Takigawa2013} is a \emph{possibility} within our model, as it is known to occur in the frustrated Ising models with anisotropic interactions~\cite{Bak.Devil}, which is what $\Hhat_{0}+\Hhat_{X}$ is, albeit with weak competing interactions ($V^{\prime}\sim V/10$ and $\Vpprime\sim V/100$ as per our estimates in Fig.~\ref{fig:Ising_Xchange}). It would be nice to see if the improved theoretical values of these effective interactions help in the occurrence of  devil's staircase. One may also consider estimating these effective parameters `phenomenologically'. 

We also like to remark that the `superfluidity' and `supersolidity' are misnomers in the context of \SrCuBorate~due to the absence of continuous symmetry in $\Hhat_{\le \frac{1}{2}}$, a quantum Ising model. The magnetisation (longitudinal or transverse) in this system does not arise by spontaneously breaking a continuous symmetry. 

Overall, this highly simple effective model presents a confident and insightful picture of the magnetisation behaviour of the SS model and \SrCuBorate, as compared to the vastly complex dimer-hard-core boson models. Our choice of the unit-cell states, it appears, is the right way to formulate and study this problem.

\begin{figure}[htbp]
\centering
\includegraphics[width=0.238\textwidth]{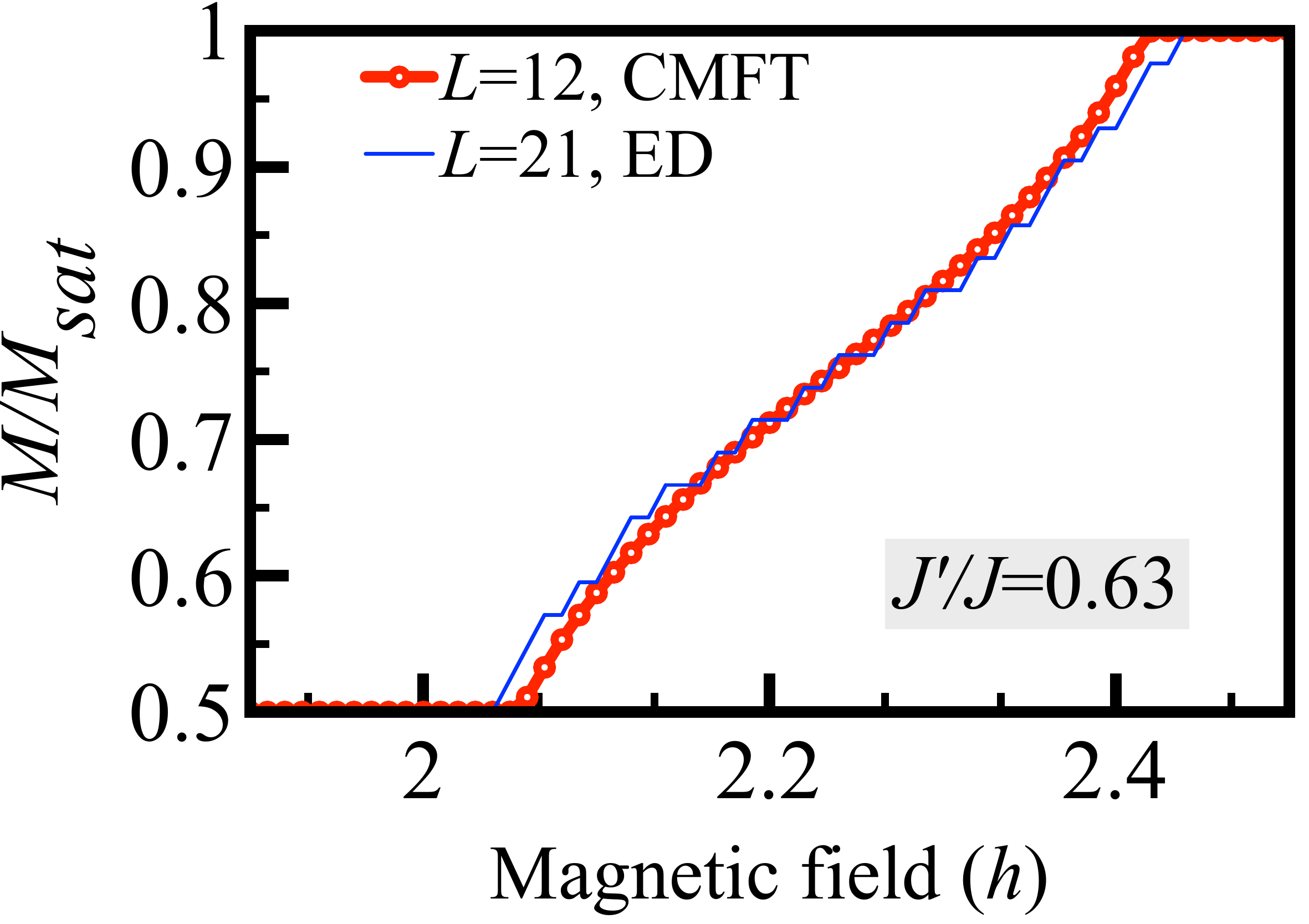}
\hfill 
\includegraphics[width=0.238\textwidth]{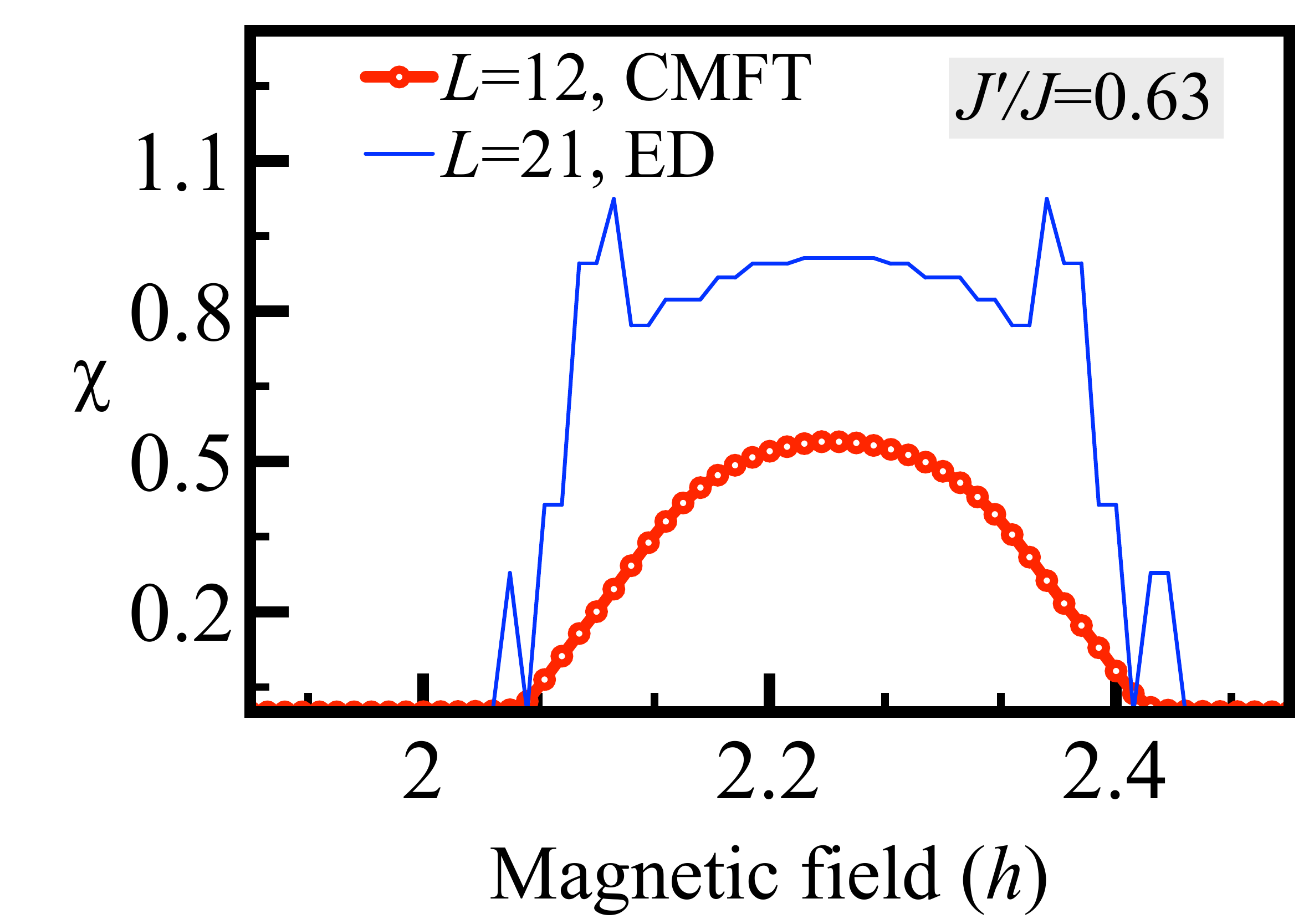}
\caption{Magnetisation and chirality ($\chi$) vs. magnetic field from the exact numerical diagonalization (ED) and cluster-mean-field theory (CMFT) calculations for $M/M_{sat}\ge 1/2$. }
\label{fig:MvsH_above1by2}
\end{figure}

%%%%%%%%%%%%%%%%%  
Finally, encouraged by the discussion below 1/2, we similarly 
derive a minimal effective Hamiltonian above the $1/2$ plateau in terms of the hard-core bosons defined in the $\{\ketOne, \ketTwo\}$ basis  (see Table~\ref{tab:Rep12}). It can be written as: 
$\Hhat_{\ge\frac{1}{2}}=-\mu\sum_{\r}\nhat_{\r} + \sum_{\r}[U \nhat_{\r}\nhat_{\r+\delta_{1}} + t(\bhat^{\dagger}_{\r} \bhat_{\r+\delta_{1}}+h.c)] +\sum_{\r}\sum_{\delta=\delta_{2},\delta_{3}} [U^{\prime} \nhat_{\r}\nhat_{\r+\delta}+t^{\prime} (\bhat^{\dagger}_{\r} \bhat_{\r+\delta}+h.c)] + e_{0}L$. The various model parameters are: $U = J^{\prime}(1-\cos\theta)(3+\cos{\theta}-2\sqrt{2}\sin{\theta})/32$, $U^{\prime}=V^{\prime}$, $t= J^{\prime}(\sqrt{2}\sin{\theta}+\cos{\theta}-1)/8$, $t^{\prime}=-J^{\prime}(1-\cos\theta)/8$, $\mu =(h-\Delta_2)+2U+4U^{\prime}+\frac{J^{\prime}}{2\sqrt{2}}(\sin{\theta}-3\sqrt{2})$, and $e_{0}=E_{(1;-)}+U +2U^{\prime}+\frac{J^{\prime}}{2\sqrt{2}}\sin{\theta}$. For $J^{\prime}/J \in [0,1]$, $t$, $U$ and $U^{\prime}$ all are positive, and $t^{\prime}<0$. Moreover, $0 \approx U \ll U^{\prime} \lesssim |t^{\prime}| < t < 0.09$. It is an XXZ model on isosceles triangular lattice, with a dominant XY part.  
Here, we again calculate the magnetisation, $M/M_{sat}=(1+\frac{1}{L}\sum_{\r}\langle \nhat_{\r}\rangle)/2$, as a function of $h$. Due to the weak $U$ and $U^{\prime}$, and  strong quantum fluctuations, we don't expect any crystalline order of triplets, and thus, no plateaus. We did exact numerical diagonalization (ED) on periodic clusters of $L$ upto 21, and a 12-sublattice cluster mean-field theory (CMFT) on a 12-sites exact cluster coupled to the mean-fields at the boundary. Both these calculations give smooth $M$ vs. $h$ curves (see Fig.~\ref{fig:MvsH_above1by2}).

%%%%%%%%%%%%
\begin{table}
\caption{\label{tab:Rep12}  Representation of the spins in a unit-cell in the basis, $\{\ketOne,\ketTwo\}$, where $\ketOne = |1,1;-\rangle$ and $\ketTwo=|2,2\rangle$.\footnote{Here, $p_{\theta} = \frac{1}{2}(\frac{1}{\sqrt{2}} \cos\frac{\theta}{2}+\frac{1}{2} \sin\frac{\theta}{2})$, $q_{\theta} = \frac{1}{2}(\frac{1}{\sqrt{2}} \cos\frac{\theta}{2}-\frac{1}{2} \sin\frac{\theta}{2})$, $\tilde{p}_{\theta}=\frac{1}{8}(1-\cos\theta-2\sqrt{2}\sin\theta)$, and $\tilde{q}_{\theta}=\frac{1}{8}(1-\cos\theta+2\sqrt{2}\sin\theta)$. Moreover, $\nhat=\ketTwo\braTwo$, $\identity=\ketOne\braOne+\ketTwo\braTwo$ and $\tau^{+}=\ketOne\braTwo\equiv \bhat^{\dag}$.}
}
\begin{ruledtabular}
\begin{tabular}{l | l | l}
$S^{x}_{1}=S^{x}_{2}=\frac{1}{4} \sin\frac{\theta}{2}\tau^{x}$ 
& $S^{x}_{3}= - p_{\theta}\tau^{x}$ & $S^{x}_{4}=q_{\theta}\tau^{x}$ \\
$S^{y}_{1}=S^{y}_{2}=-\frac{1}{4} \sin\frac{\theta}{2}\tau^{y}$ 
& $S^{y}_{3}= p_{\theta}\tau^{y}$ & $S^{y}_{4}=-q_{\theta}\tau^{y}$ \\
$S^{z}_{1}=S^{z}_{2}=\frac{3+\cos\theta}{8}(\identity-\nhat) $
& $S^{z}_{3}=\tilde{p}_{\theta}(\identity-\nhat) $ & 
$S^{z}_{4}=\tilde{q}_{\theta}(\identity-\nhat) $ \\
\hspace{.13\textwidth} $+\frac{1}{2}\nhat$ & 
\hspace{.05\textwidth} $+\frac{1}{2}\nhat$ & \hspace{.05\textwidth} $+\frac{1}{2}\nhat$
\end{tabular}
\end{ruledtabular}
\end{table}
%%%%%%%%%%%%%%%%%%%

Since the XY model on triangular lattice exhibits chiral order~\cite{Suzuki1997}, we also calculate it in the ground state of $\Hhat_{\ge\frac{1}{2}}$ as a function of $h$. The $z$-component of chirality of an upright triangle at position $\R$ is written as, $\chi^z(\R) = (\vec\tau_1\times \vec\tau_2)_z+ (\vec\tau_2\times \vec\tau_3)_z+(\vec\tau_3\times \vec\tau_1)_z$, where 1, 2 and 3 are the spins of that triangle. The chiral order parameter is defined as: $\chi = \sqrt{[\sum_{\R} \chi^z(\R)]^2/LS(LS+1)}$, where $S=1/2$ and $\R$ runs over the upright triangles of a cluster~\cite{Suzuki1997}. The data in Fig.~\ref{fig:MvsH_above1by2} clearly indicates the presence of chiral order for $M/M_{sat}$ between $1/2$ and 1, while the plateaus are absent. Close to 1/2 and 1, however, the spikiness in $\chi$ seem to indicate some anomalies that may show up in $M$ (possibly as jumps).

\begin{acknowledgments}
We acknowledge DST-FIST support for the computational facilities. B. D. thanks CSIR for financial support.
\end{acknowledgments}
%%%%%%%%%%%
\bibliography{manuscript.bib}
\end{document}